\documentclass[11pt,a4paper, final, twoside]{article}
\usepackage{amsmath}
\usepackage{fancyhdr}
\usepackage{amsthm}
\usepackage{amsfonts}
\usepackage{amssymb}
\usepackage{amscd}
\usepackage{latexsym}
\usepackage{amsthm}
\usepackage{graphicx}
\usepackage{graphics}
\usepackage{afterpage}
\usepackage[colorlinks=true, urlcolor=blue,  linkcolor=black, citecolor=black]{hyperref}
\usepackage{color}
\theoremstyle{definition}

\theoremstyle{remark}

\numberwithin{equation}{section}
\usepackage{float}
\usepackage[dvipsnames]{xcolor}
\usepackage{bm}

\usepackage{textcomp}
\usepackage{gensymb}

\begin{document}

\renewcommand{\headrulewidth}{1pt}
\newcommand{\HRule}{\rule{\linewidth}{1pt
}}

\hyphenpenalty=100000
\begin{flushright}
{\Large \textbf{\\The Differential Equations of Gravity-free Double Pendulum: Lauricella Hypergeometric Solutions and Their Inversion.}}\\[5mm]
{\large \textbf{Alessio Bocci $^\mathrm{1^*}$\footnote{\emph{*Corresponding author: E-mail: alessiobocci@live.it}} and Giovanni Mingari Scarpello $^\mathrm{2}$ }}\\[3mm]
$^\mathrm{1}${\footnotesize \it Department of Aerospace Science and Technology, Politecnico di Milano, Italy.\\ 
$^\mathrm{2}$ Ordine Degli Ingegneri di Milano, Italy.}
\end{flushright}
\begin{flushleft}\fbox{%
\begin{minipage}{1.3in}
{\slshape \textbf{Original Research Article}\/}
\end{minipage}}
\end{flushleft} 
\HRule\\[3mm]
{\Large \textbf{Abstract}}\\[4mm]
\fbox{%
\begin{minipage}{5.4in}{\footnotesize 
This paper solves in closed form the system of ODEs ruling the 2D motion  of a gravity free double pendulum (GFDP), not subjected to any force. In such a way its movement is governed by the initial conditions only. 
   The relevant strongly non linear ODEs have been put back to hyperelliptic quadratures which, through the Integral Representation Theorem (IRT), are driven to the Lauricella hypergeometric functions.

 We compute time laws and trajectories of both point masses forming the GFDP in explicit closed form. Suitable sample problems are carried out in order to prove the method effectiveness.
} \end{minipage}}\\[4mm]


\section{Introduction}

The system consists of two point-masses $m_1$ and $m_2$ moving over an horizontal fixed plane, where the distance between a fixed point $P$ (called pivot) and $m_1$ and the distance between $m_1$ and $m_2$ are fixed and equal to $l_1$ and $l_2$ respectively: we can think to a couple of massless beams. In absence of gravity and any other external force, the motion is driven by the four initial conditions only. Assuming the pivot as a pole and as a polar axis the horizontal line passing through it, the planar line $\Gamma$ referred to such a frame will denote the $m_2$ absolute trajectory given in polar coordinates $\rho$ and $\mu$.

\begin{figure}
    \centering
    \includegraphics[scale=1.1]{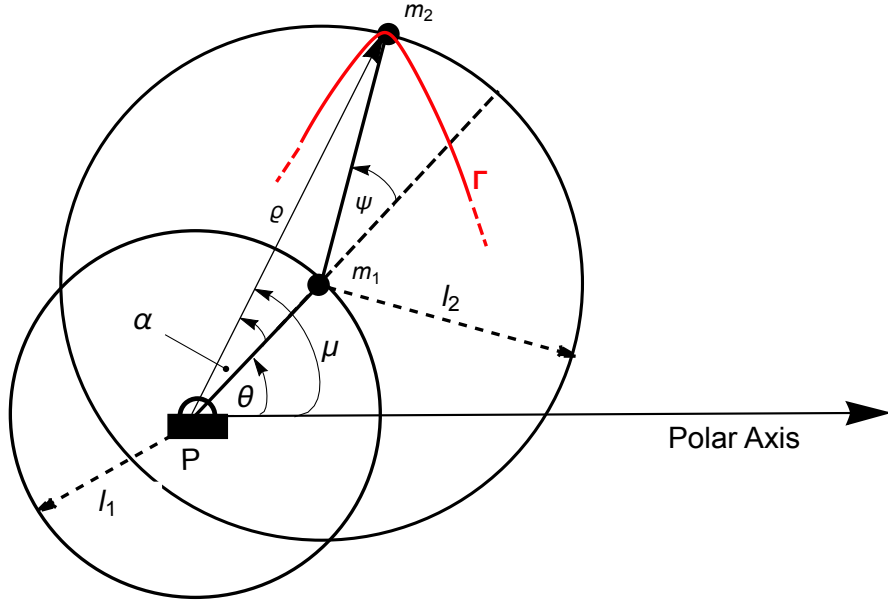}
    \caption{GFDP circles of motion and $\Gamma$ trajectory of $m_2$.}
    \label{fig1}
\end{figure}

The selected lagrangian coordinates are the angle ${\theta}$ (between the polar axis and the first beam), and ${\psi}$ ( the angle between the first and second beam) both positive  counterclockwise.

Within the nonlinear systems there is a hierarchy of complexity: the simple pendulum of small amplitude has harmonic oscillations and a period independent from the amplitude; 
in the case of large amplitudes, such dependence occurs, but the regularity of motion holds.
The heavy double pendulum  in the case of large amplitudes changes radically with increasing energy and the oscillations become chaotic.

 In \cite{Richter,Szuminski}   a general integrability study about the double pendulum is presented. In particular in \cite{Szuminski}  the integrability of GFDP is assessed (i.e. it has $\mathbb{S}^1$ symmetry, and the Lagrange function depends on difference of angles only) and also more complicated systems (e.g. GFDP with multiple springs) are  analyzed by mean of Poincar\'{e} sections. In \cite{Richter} the GFDP problem is also led to the quadratures, but the resulting hyperelliptic integrals are not solved.  
The so called Melnikov method is often used to predict the occurrence of chaotic orbits in non-autonomous nonlinear systems; in effect it appeared in 1890 by H. Poincar\'{e}, but was successfully displayed by V. Melnikov in 1963. Anyway, according to it, one constructs a \emph{Melnikov function}, which determines a measure of the distance between stable and unstable manifolds in the Poincar\'{e} map. When this measure is by the method found to be zero, those manifolds cross each other transversally and the system behavior will become chaotic.
Melnikov's method has been applied by Dullin \cite{Dullin} to the planar double (heavy!) pendulum proving it -generally speaking- is a chaotic system. The parameter space of the double pendulum is by him discussed, and two kinds of integrable cases are identified. The first is just without gravity: we have two dynamically coupled rotators with a stable orbit in the stretched configuration and an unstable orbit in the folded configuration. This integrable limit case, namely the GFDP, can only be solved in terms of hyperelliptic functions \cite{Richter} . 
 Dullin studies also the other integrable case of the double pendulum, where the system consists of two uncoupled pendulums. The main analytical result for the GFDP problem is provided by Enolsky \cite{enolskii2003double} employing an analytical approach, fully different from ours, founded on Theta Riemann series \cite{fay2006theta}.  

 As far as we are concerned, the literature on the GFDP non linear ODE system does not appear to have further items.

Let us now do an overview of the main pros and cons of our approach.
First, all solutions are formally given in closed form requiring a numerical support only restricted to the Fourier series expansion coefficients, whose only few  terms are necessary for an acceptable accuracy. Second, being the solution provided in terms of sines and cosines, it can easy handled for each computational necessity. But the highest score is the \emph{inversion}, namely to have finally obtained the motion time laws in explicit way.
Coming to the cons aspects, we based on Lauricella functions which are not available in software packages. We then used their series expansion based on some our special equivalence relationship we will describe in a next paper.
The problem of GFDP is of high interest in the field of engineering and robotics (some advanced techniques are presented in \cite{talamucci2018synchronization}) and our zero gravity treatment can really be a first guess for more complicated variants when the weight can be neglected (i. e. micro-gravity environments).
Despite to its practice implications, the literature on it is quite poor, specially as it concerns the analytical treatments.

\section{List of symbols}

    \begin{tabular}{ll}
        $A$ & motion first integral   \\
       $a_0$, $a_n$, $b_n$  & Fourier expansion coefficients\\
       $B$ &constant value of the Hamiltonian \\
       $F_D$ & Lauricella hypergeometric function \\
       $H$ & Hamilton function \\
       $L$ & Lagrange function \\
       $l_1$ & fixed distance between the first bob and pivot\\
       $l_2$ &fixed distance between the second and first bob\\ 
       $m_1$ & mass of the bob connected to the pivot \\
       $m_2$ & mass of the second bob \\
       $n$ & nth. harmonic of the Fourier series expansion\\ 
       $P$ & pivot and reference pole \\
       $p_\theta$ & conjugate moment with reference to $\theta$\\
       $t$ & generic time instant\\
       $t_0$ & generic initial instant \\
       $\hat{t}$, $\tilde{t}$ & extrema time instants identified by $w$-function\\
       $T_i$ & different periods of motion \\
       $y$ & $=\cos(\psi)$\\
       $\alpha$ & $=\mu-\theta$\\
       $\beta$ & ratio between lengths $l_1$ and $l_2$\\
       $\gamma$ & $=(\nu+1)\beta^2+1$\\
       $\Gamma$& planar curve described by $m_2$\\
       $\delta$ & $=\sqrt{\nu+1}$\\
       $\epsilon$ & non dimensional system parameter \\
       $\zeta$ & $=\gamma/(2\beta)$\\
       $\theta$ & polar anomaly of bob $m_1$\\
       $\theta_0$ & initial condition on $\theta$ variable \\
       $\dot{\theta}_0$ & initial  $\theta$-velocity \\
       $\lambda$ & auxiliary variable $=y-\epsilon$\\
       $\mu$ & polar anomaly of bob $m_2$\\
       $\nu$ & ratio of masses $m_1$ and $m_2$\\
       $\xi_n$ & $=2\pi n/T$\\
       $\rho$ & polar radius of $m_2$\\
       $\sigma$ & $=(1-|\epsilon|)/2$\\
       $\phi$ & polar anomaly of $\overline{m_1m_2}$\\
       $\psi=\phi-\theta$ & relative inclination of $\overline{m_1m_2}$\\
       $\psi_0$ & initial condition on $\psi$ variable\\
       $\dot{\psi}_0$ & initial  $\psi$-velocity 
    \end{tabular}

\section{The motion differential equations}
Since no potential field of energy is present,  Lagrangian ($L$) and the Hamiltonian ($H$) functions will coincide with kinetic energy:

\begin{equation*}
L=H=\frac{1}{2}m_1 l_1^2\dot{\theta}^2+\frac{1}{2} m_2 \left[l_1^2 \dot{\theta}^2+l_2^2 \dot{\phi}^2+2 l_1 l_2 \cos(\phi-\theta)\dot{\phi}\dot{\theta} \right]
\end{equation*}

Setting $\psi=\phi-\theta$, $\nu=m_1/m_2$, $\beta=l_1/l_2$ and $\gamma=(\nu+1)\beta^2+1$ we get:

\begin{equation*}
 L=H= \frac{1}{2} m_2 l_2^2  \left[\gamma
  +2 \beta  \cos
   (\psi )\right]\dot{\theta}
  ^2+m_2 l_2^2 
   [1+\beta  \cos (\psi
   )]\dot{\theta}\dot{  \psi }+ \frac{1}{2} m_2 l_2^2\dot{\psi} ^2
\end{equation*}

The conjugate moment relevant to the coordinate $\theta$ will be a motion first integral, say $A$:

\begin{equation*}
\frac{p_\theta}{m_2l_2^2}=\frac{1}{m_2l_2^2}\frac{\partial L}{\partial \dot{\theta}}=\left[\gamma
  +2 \beta  \cos
   (\psi )\right]\dot{\theta}+[1+\beta  \cos (\psi
   )]\dot{\psi}=A
\end{equation*}

Since no dissipation occurs, then the hamiltonian H
is also constant, say $B$;  the coupled  differential equations system is:

 \begin{equation}
 \begin{cases}
 \left[\gamma
  +2 \beta  \cos
   (\psi )\right]\dot{\theta}+[1+\beta  \cos (\psi)]\dot{\psi}=A & A\in \mathbb{R}\\
   \left[\gamma
  +2 \beta  \cos
   (\psi )\right]\dot{\theta}
  ^2+2[1+\beta  \cos (\psi
   )]\dot{\theta}\dot{  \psi }+ \dot{\psi} ^2=B & B\in \mathbb{R}
 \end{cases}
 \label{sys}
 \end{equation}
 
Solving \eqref{sys} for $\dot{\psi}$ and $\dot{\theta}$ we obtain the differential equations of motion. The first is:

\begin{equation}\label{finalmente}
\begin{cases}
\dot{\psi}= 
   \mp \sqrt{\dfrac{ A^2-2
   \beta  B \cos (\psi )-B \gamma }{\beta ^2 \cos ^2(\psi )-\gamma +1}} \\
   \psi(t_0)=\psi_0
\end{cases}
\end{equation}


$$
$$ 

and:
\begin{equation}\label{lalla}
\begin{cases}
    \dot{\theta}=\pm \dfrac{\beta \cos(\psi)+1}{2\beta \cos(\psi)+\gamma} \dot{\psi} \mp \dfrac{A}{2\beta \cos(\psi)+\gamma}\\
    \theta(t_0)=\theta_0
    \end{cases}
\end{equation}

\subsection{The \texorpdfstring{$\psi$}{TEXT} solution} 
Performing the change of variable:
\begin{equation}
\cos(\psi)=y \rightarrow d\psi=-\frac{1}{\sqrt{1-y^2}} dy
\label{cv}
\end{equation}

then \eqref{finalmente} becomes:

\begin{equation}
\begin{aligned}
t-t_0 &=
\pm \sqrt{\frac{\beta}{2B}} \int_{y_0}^y \sqrt{\frac{(s-\delta)(s+\delta)}{\left(\epsilon-s\right)(1+s)(1-s)}}{\rm d}s= \pm \sqrt{\frac{\beta}{2B}}\int_{y_0}^y f(s){\rm d}s
\end{aligned}
\label{su}
\end{equation}

with:
\begin{equation}
    \epsilon=\dfrac{A^2-B \gamma }{2 \beta  B}, \quad \delta=\sqrt{\nu+1}>1, \quad  f(s)=\sqrt{\frac{(s-\delta)(s+\delta)}{(\epsilon-s)(1+s)(1-s)}}
\end{equation}

\subsection{The \texorpdfstring{ $\theta$}{TEXT} solution}

Dividing \eqref{lalla} to \eqref{finalmente} and making the same change of variable \eqref{cv}, we obtain $\theta$ as a function of $y=\cos\psi$:

\begin{equation*}
    \theta-\theta_0= I_1(y)+I_2(y)
\end{equation*}

where:

\begin{align*}
    I_1(y)&=\mp \frac{1}{2}\int_{y_0}^y \frac{s+1/\beta}{(  s+ \zeta)\sqrt{1-s^2}}{\rm d}s=\\
    &=\mp \frac{1}{2}\left[\sin^{-1}(s)+\frac{1/\beta-\zeta}{\sqrt{\zeta^2-1}} \tan ^{-1}\left(\frac{ \zeta s+1}{\sqrt{(1-s^2)(\zeta^2-1)}}\right)\right]_{y_0}^y
\end{align*}

with:

\begin{equation*}
 \zeta= \frac{\gamma}{2\beta}>1
\end{equation*}
Notice that $\zeta >1$ always; indeed the contrary would lead to:

\begin{equation*}
    (\nu+1)\beta^2 -2 \beta+1<0 
\end{equation*}

which cannot be true.

In order to complete the function $\theta$ computing, the further integral has to be carried out:

\begin{equation}
\begin{aligned}
I_2(y)&=
\pm \frac{A }{2\sqrt{2B\beta}}\int_{y_0}^y \frac{f(s)}{s+\zeta}{\rm d}s
\end{aligned}
\label{sa}
\end{equation}

Our main trouble is due to its upper limit $y$, namely the inverse of the hyperelliptic integral \eqref{su} which therefore becomes our next step.

\section{ The three allowed motions}





The highest possible span of $y$ is of course $y\in [-1,1]$; nevertheless it will be narrowed in some cases. Furthermore, by its definition $\delta >1$: then these roots lie outside the $y$-domain. As a consequence, the only root really influencing all the features of the whole GFDP motion is the non dimensional parameter $\epsilon$.

We see 4 cases at all:
\begin{enumerate}
\item 
If  $0 \leq \epsilon \leq 1$ then: $ y\in [\epsilon,1]$, so that the angle $\psi$ cannot assume all the possible values. Anyway $y$ is time-periodical and its period will be given by:
\begin{equation*}
T_{1}=\frac{2\beta}{\sqrt{2B\beta}} \int_{\epsilon}^1 f(s){\rm d}s
\end{equation*}
\item 
If $-1\leq \epsilon \leq 0$ then $y \in [-|\epsilon|,1]$, namely the angle $\psi$ has some different restrictions and the $y$ time-period will be:
\begin{equation*}
T_{2}=\frac{2\beta}{\sqrt{2B\beta}} \int_{-|\epsilon|}^1 f(s){\rm d}s
\end{equation*}
\item 
If $\epsilon<-1$ then $y\in [-1,1]$ and this is the only occurrence where the angle $\psi$ is free of taking all the possible values.
The $y(t)$ time-period will be: 
\begin{equation*}
T_{3}=\frac{2\beta}{\sqrt{2B\beta}} \int_{-1}^1 f(s){\rm d}s
\end{equation*}
\item 
If $\epsilon >1$ the motion cannot exist, being the argument of the square root in \eqref{finalmente} less than zero  $ \forall y \in [-1,1]$. 
\end{enumerate}

This analysis shows $\epsilon$ as the crucial parameter ruling the whole system  behavior. 




\section{First motion: \texorpdfstring{ $0 < \epsilon < 1$}{TEXT}}
\subsection{Integration}
In such a case $ y\in [\epsilon,1]$ and we have: 



\begin{equation}
    t-t_0= \pm\sqrt{\frac{\beta}{2B}}\int_{y_0}^y f(s){\rm d}s, \quad \text{with} \quad y \in ]\epsilon,1[ , \quad \epsilon \in ]0,1[
    \label{uno}
\end{equation}

in order to solve \eqref{uno} with $\epsilon \in [0,1]$ in closed form, let us pass from $s$ to $u$ through a first change of variable:

\begin{equation*}
    s=u+\epsilon
\end{equation*}

i.e. we move to $\epsilon$ the origin of the variable of integration; let us now define:

\begin{equation*}
    g(u)=f(u+\epsilon)=\sqrt{\frac{(q_1-u)(q_2+u)}{u(q_3-u)(q_4+u)}}
\end{equation*}

with:

\begin{equation*}
    q_1=\delta-\epsilon, \quad q_2=\delta+\epsilon, \quad q_3=1-\epsilon, \quad q_4=1+\epsilon
\end{equation*}
So that:
\begin{equation*}
\begin{aligned}
    &\int_{y_0}^y f(s){\rm d}s= \int_{y_0-\epsilon}^{y-\epsilon} g(u){\rm d}u= \\ &=\int_0^{y-\epsilon} g(u){\rm d}u-\int_{0}^{y_0-\epsilon}g(u){\rm d}u=G(y-\epsilon)-G(y_0-\epsilon)
\end{aligned}    
\end{equation*}

where:

\begin{equation}
    G(\lambda)=\int_0^\lambda g(u){\rm d}u, \quad \text{with} \quad \lambda=y-\epsilon, \quad \text{so that:}\quad \lambda \in ]0,1-\epsilon[
\label{due}
\end{equation}

In order to solve \eqref{due}, let us resort to a further change from $u$ to $v$:

\begin{equation*}
    u=\lambda v
\end{equation*}

leading to:

\begin{equation*}
G(\lambda)=\lambda \int_0^1 g(\lambda v) dv     
\end{equation*}

After some algebraic manipulations we arrive at:


\begin{equation}
G(\lambda)=\sqrt{\lambda \hat{q}} \int_0^1 v^{-\frac{1}{2}} \left(1-\hat{\lambda}_1 v\right)^{-\frac{1}{2}}
\left(1-\hat{\lambda}_2 v\right)^{\frac{1}{2}}
\left(1- \hat{\lambda}_3 v\right)^{-\frac{1}{2}}   
\left(1-  \hat{\lambda}_4 v\right)^{\frac{1}{2}}   {\rm d}v
\label{dio}
\end{equation}

with:
\begin{equation*}
    \hat{\lambda}_1=\frac{\lambda}{q_3}, \quad \hat{\lambda}_2=\frac{\lambda}{q_1}, \quad \hat{\lambda}_3=-\frac{\lambda}{q_4}, \quad
    \hat{\lambda}_4=-\frac{\lambda}{q_2}, \quad
    \hat{q}=\frac{q_1 q_2}{q_3q_4}
\end{equation*}

The reader \eqref{dio} is referred  to the IRT \eqref{iirtt} for the Lauricella hypergeometric functions. We are faced to a $F_D^{(4)}$: 


\begin{equation}
    G(\lambda)=2\sqrt{\frac{\lambda q_1 q_2}{q_3 q_4}} F_D^{(4)} \left(\left.\begin{array}{c}\begin{array}{cc}
      \dfrac{1}{2};   & \dfrac{1}{2},-\dfrac{1}{2},\dfrac{1}{2},-\dfrac{1}{2}\end{array}\\
      \\
       \dfrac{3}{2} 
    \end{array} \right| \dfrac{\lambda}{q_3},\dfrac{\lambda}{q_1},-\dfrac{\lambda}{q_4},-\dfrac{\lambda}{q_2}\right)
    \label{tre}
\end{equation}

Many properties of hypergeometric functions can be found in \cite{Koepf}. By \eqref{tre} one can check that Lauricella's arguments are-as required by the representation theorem- less than unity. So that, going back, we could infer that:

\begin{equation}
  t-t_0= \pm \left[
    \eta(s) F_D^{(4)} \left(\left.\begin{array}{c}\begin{array}{cc}
      \dfrac{1}{2};   & \dfrac{1}{2},-\dfrac{1}{2},\dfrac{1}{2},-\dfrac{1}{2}\end{array}\\
      \\
       \dfrac{3}{2} 
    \end{array} \right| \dfrac{s-\epsilon}{1-\epsilon},\dfrac{s-\epsilon}{\delta-\epsilon},\dfrac{\epsilon-s}{1+\epsilon},\dfrac{\epsilon-s}{\delta+\epsilon}\right)\right]_{y_0}^y
    \label{tempo}
\end{equation}
where:
\begin{equation*}
    \eta(s)=\sqrt{\frac{2\beta(s-\epsilon) (\delta^2-\epsilon^2) }{B(1-\epsilon^2)}} 
\end{equation*}

What above provides time as a hypergeometric function of four variable-ratios all related to $y$ and then to $\psi$: this is the required $t=t(y)$.
But \eqref{sa} requires $y$, so that the last outcome has to be inverted.
\subsection{ Period evaluation of \texorpdfstring{$y(t)$}{TEXT}}

Of course the period $T_1$ could be carried out by means of \eqref{tre}, namely $  F_D^{(4)}$, nevertheless we preferred the following path employing  $F_D^{(3)}$ which is simpler and faster.
We have:

\begin{equation*}
    \lambda=1-\epsilon=q_3
\end{equation*}

By means of \eqref{dio} we get:

\begin{equation*}
    T_{1}=\pi \sqrt{\frac{2\beta(\delta^2-\epsilon^2)}{B(1+\epsilon)}} F_D^{(3)} \left(\left.\begin{array}{c}\begin{array}{cc}
      \dfrac{1}{2};   & -\dfrac{1}{2},\dfrac{1}{2},-\dfrac{1}{2}\end{array}\\
      \\
      1 
    \end{array} \right| \dfrac{1-\epsilon}{\delta-\epsilon},\dfrac{\epsilon-1}{1+\epsilon},\dfrac{\epsilon-1}{\delta+\epsilon}\right)
\end{equation*}

\subsection{Fourier coefficients of \texorpdfstring{$y(t)$}{TEXT} expansion}

The Fourier-approach followed hereinafter has been successfully tested  in our previous paper \cite{bocci2021analytic} and some similar applications are presented in \cite{gimeno2009rational}.

In order to invert the $t(y)$  \eqref{tempo}, let us refer to $y(t)$ Fourier expansion of the form:
\begin{equation*}
    y(t)=\frac{a_0}{2}+\sum_{n=1}^{+\infty} \left[{{a}_n}{\cos \left(\xi_n t\right) }+{{b}_n}{\sin \left(\xi_n t\right)}\right], \quad \xi_n= \frac{2\pi n}{T}
    \end{equation*}
where $T$ is the period of the function to be expanded. 

The a.m. coefficients have to be computed as:
$$             
a_0=\frac{2}{T} \int_ {0}^ {T}  y(t){\rm d}t, \quad
a_n=\frac{2}{T} \int_ {0}^ {T}  y(t)\cos \left(\xi_n t\right){\rm d}t, \quad
b_n=\frac{2}{T} \int_ {0}^ {T}  y(t)\sin \left(\xi_n t\right){\rm d}t
$$

Let us define the function:

\begin{equation*}
\hat{f}(s)=\sqrt{\frac{\beta}{2B}} f(s)
\end{equation*}

If $\epsilon>1/2$ then:
\begin{align*}
    a_0&=\frac{4}{T_{1}}\int_{\epsilon}^1 s \hat{f}(s)ds=\\
    &=\frac{4\pi \epsilon}{T_{1}}\sqrt{\frac{\beta(\delta^2-\epsilon^2)}{2B(1+\epsilon)}}F_D^{(4)} \left(\left.\begin{array}{c}\begin{array}{cc}
      \dfrac{1}{2};   & -\dfrac{1}{2},\dfrac{1}{2},-\dfrac{1}{2},-1\end{array}\\
      \\
      1 
    \end{array} \right| \dfrac{1-\epsilon}{\delta-\epsilon},\dfrac{\epsilon-1}{1+\epsilon},\dfrac{\epsilon-1}{\delta+\epsilon},\dfrac{\epsilon-1}{\epsilon}\right)
\end{align*}

To compute $a_0$, we do the change:
\begin{equation*}
    s=u+\sigma, \quad \sigma=\frac{1+\epsilon}{2} 
\end{equation*}
Proceeding as previously shown we have:

\begin{equation*}
    a_0= \frac{8\sigma}{T}\sqrt{\frac{\beta(\delta^2-\sigma^2)}{2B(\sigma+1)}} \sum_{i=1}^2 F_D^{(5)} \left(\left.\begin{array}{c}\begin{array}{cc}
      1;   & -1,-\dfrac{1}{2},-\dfrac{1}{2},\dfrac{1}{2},\dfrac{1}{2}\end{array}\\
      \\
      \dfrac{3}{2} 
    \end{array} \right| \hat{\sigma}_{1,i},\hat{\sigma}_{2,i},\hat{\sigma}_{3,i},\hat{\sigma}_{4,i},-1\right)
\end{equation*}
with:
\begin{equation*}
 \hat{\sigma}_{1,i}=(-1)^i\dfrac{1-\sigma}{\sigma}, \quad \hat{\sigma}_{2,i}=(-1)^{i+1}\dfrac{1-\sigma}{\delta-\sigma}, \quad \hat{\sigma}_{3,i}=(-1)^i\dfrac{1-\sigma}{\delta+\sigma}, \quad \hat{\sigma}_{4,i}=(-1)^i\dfrac{1-\sigma}{1+\sigma}
\end{equation*}

Now we define the function:

\begin{equation}
    w(y)=\eta(y) F_D^{(4)} \left(\left.\begin{array}{c}\begin{array}{cc}
      \dfrac{1}{2};   & \dfrac{1}{2},-\dfrac{1}{2},\dfrac{1}{2},-\dfrac{1}{2}\end{array}\\
      \\
       \dfrac{3}{2} 
    \end{array} \right| \dfrac{y-\epsilon}{1-\epsilon},\dfrac{y-\epsilon}{\delta-\epsilon},\dfrac{\epsilon-y}{1+\epsilon},\dfrac{\epsilon-y}{\delta+\epsilon}\right)
    \label{w(s)}
\end{equation}

Let it be:
\begin{equation*}
    \hat{t}=w(y_0), \quad y_0=\cos(\psi_0) \quad \text{and} \quad \dot{y}_0=-\sin(\psi_0) \dot{\psi}_0<0
\end{equation*}

Considering a period, the piece-wise defined function which describes the $t=t(y)$ is:

\begin{equation*}
t(y)=
    \begin{cases}
    t_1(y)= \hat{t}-w(y) & y \in [y_0,\epsilon]\\
    t_2(y)=\hat{t}+w(y) & y\in[\epsilon,1]\\
    t_3(y)=\hat{t}+T-w(y) & y\in [1,y_0]
    \end{cases}
\end{equation*}

Marking as $t'$ the prime derivative of $t$ with respect to $y$, we get:

\begin{equation*}
    t'(y)=
    \begin{cases}
    t_1'(y)=-\hat{f}(y)  & y \in [y_0,\epsilon]\\
    t_2'(y)=\hat{f}(y) & y \in [\epsilon,1] \\
    t_3'(y)=-\hat{f}(y) &  y\in [1,y_0]
    \end{cases}
\end{equation*}

Defining $\hat{a}=T a_n/2$, we have:

\begin{align*}
\hat{a}_n &= \int_0^T y(t) \cos (\xi_n t)dt=-\int_{y_0}^\epsilon s \hat{f}(s)\cos(\xi_n t_1(s)){\rm d}s+ \\
&+\int_{\epsilon}^1 s \hat{f}(s)\cos(\xi_n t_2(s))ds-\int_{1}^{y_0}s\hat{f}(s) \cos(\xi_n t_3(s)){\rm d}s
\end{align*}

where has been done the change $t=y^{-1}(s)$ and then ${\rm d}t=t'(s){\rm d}s$.

Evaluating apart each of the three terms of the above formula, we get:

\begin{align*}
    -\int_{y_0}^\epsilon s \hat{f}(s)\cos(\xi_n t_1(s))ds&=\left[ \frac{s}{\xi_n} \sin(\xi_n t_1(s))\right]_{y_0}^{\epsilon}-\frac{1}{\xi_n}\int_{y_0}^{\epsilon}\sin(\xi_n t_1(s)){\rm d}s\\
    +\int_{\epsilon}^1 s \hat{f}(s)\cos(\xi_n t_2(s)){\rm d}s &=\left[ \frac{s}{\xi_n} \sin(\xi_n t_2(s))\right]_{\epsilon}^{1}-\frac{1}{\xi_n}\int_{\epsilon}^{1}\sin(\xi_n t_2(s)){\rm d}s\\
    -\int_{1}^{y_0}s \hat{f}(s) \cos(\xi_n t_3(s)){\rm d}s &=\left[ \frac{s}{\xi_n} \sin(\xi_n t_3(s))\right]_{1}^{y_0}-\frac{1}{\xi_n}\int_{1}^{y_0}\sin(\xi_n t_3(s)){\rm d}s
\end{align*}

It is easy to see that adding the three equations, the sum of the square brackets is zero. Furthermore:

\begin{align*}
    \sin(\xi_n t_1(s))&=\sin(\xi_n \hat{t})\cos(\xi_n w(s))-\cos(\xi_n \hat{t})\sin(\xi_n w(s))\\
    \sin(\xi_n t_2(s))&=\sin(\xi_n \hat{t})\cos(\xi_n w(s))+\cos(\xi_n \hat{t})\sin(\xi_n w(s))\\
    \sin(\xi_n t_3(s))&=\sin(\xi_n (\hat{t}+T))\cos(\xi_n w(s))-\cos(\xi_n( \hat{t}+T))\sin(\xi_n w(s))
\end{align*}

But:

\begin{align*}
    \sin(\xi_n(\hat{t}+T))=\sin(\xi_n \hat{t})\cos(\xi_n T)+\cos(\xi_n \hat{t})\sin(\xi_n T)=\sin(\xi_n \hat{t})\\
    \cos(\xi_n(\hat{t}+T))=\cos(\xi_n \hat{t})\cos(\xi_n T)-\sin(\xi_n \hat{t})\sin(\xi_n T)=\cos(\xi_n \hat{t})
\end{align*}

so that:

\begin{equation*}
    \sin(\xi_n t_3(s))=\sin(\xi_n \hat{t})\cos(\xi_n w(s))-\cos(\xi_n \hat{t})\sin(\xi_n w(s))
\end{equation*}

By substitution, we obtain:

\begin{equation}
    \hat{a}_n=-2 \frac{\cos(\xi_n\hat{t})}{\xi_n}\int_{\epsilon}^1\sin(\xi_n w(s)){\rm d}s
    \label{anspec}
\end{equation}

Minding that $w(s)$ is the hypergeometric Lauricella function \eqref{w(s)}, the previous formula provides a {\it numerical} recipe in order to compute the $a_n$ coefficients as required.
After defining $\hat{b}_n=Tb_n/2$, an analogous procedure leads to:

\begin{equation*}
    \hat{b}_n=-2 \frac{\sin(\xi_n\hat{t})}{\xi_n}\int_{\epsilon}^1\sin(\xi_n w(s)){\rm d}s
\end{equation*}

so that a link between the coefficients is found as:

\begin{equation*}
    \hat{b}_n=\tan(\xi_n \hat{t})\hat{a}_n
\end{equation*}

The expansion of  $y(t)$ in Fourier series therefore results to be:

\begin{equation*}
    y(t)=\frac{a_0}{2}+\frac{2}{T}\sum_{n=1}^{+\infty} \frac{\hat{a}_n}{\cos(\xi_n \hat{t})}\cos(\xi_n(t-\hat{t}))
\end{equation*}
In such a way $\cos\psi=y$ has been found as a function of time.

\subsection{Detection of \texorpdfstring{$\theta(t)$}{TEXT}}
By \eqref{sa} we can evaluate the integral $I_2(y)$
We get:

\begin{equation*}
   I_2(y)= \left[\tilde{\eta}(s) F_D^{(5)} \left(\left.\begin{array}{c}\begin{array}{cc}
      \dfrac{1}{2};   & -\dfrac{1}{2},-\dfrac{1}{2},1,\dfrac{1}{2},\dfrac{1}{2}\end{array}\\
      \\
       \dfrac{3}{2} 
    \end{array} \right| \dfrac{\epsilon-s}{\delta+\epsilon},\dfrac{s-\epsilon}{\delta-\epsilon},\dfrac{\epsilon-s}{\zeta+\epsilon},\dfrac{\epsilon-s}{1+\epsilon},\frac{s-\epsilon}{1-\epsilon}\right)\right]_{y_0}^y
\end{equation*}

where:

\begin{equation*}
    \tilde{\eta}(s)=\frac{A}{\zeta+\epsilon}\sqrt{\frac{(s-\epsilon)(\delta^2-\epsilon^2)}{2\beta B (1-\epsilon^2)}}
\end{equation*}
In such a way $\theta$ has been computed as a function of time.

\subsection{Sample problem, first motion }

For the sample problem we assume:

\begin{equation*}
    \nu=\frac{1}{2}, \quad \beta=\frac{1}{2}, \quad \psi_0=70 \degree, \quad \theta_0=22.5 \degree, \quad \dot{\psi}_0=2 \quad \frac{\text{rad}}{\text{s}}, \quad \dot{\theta}_0=1 \quad\frac{\text{rad}}{\text{s}}
\end{equation*}

and therefrom we have $\epsilon=0.209051$ .

\begin{figure}[H]
    \centering
    \includegraphics[scale=1.25]{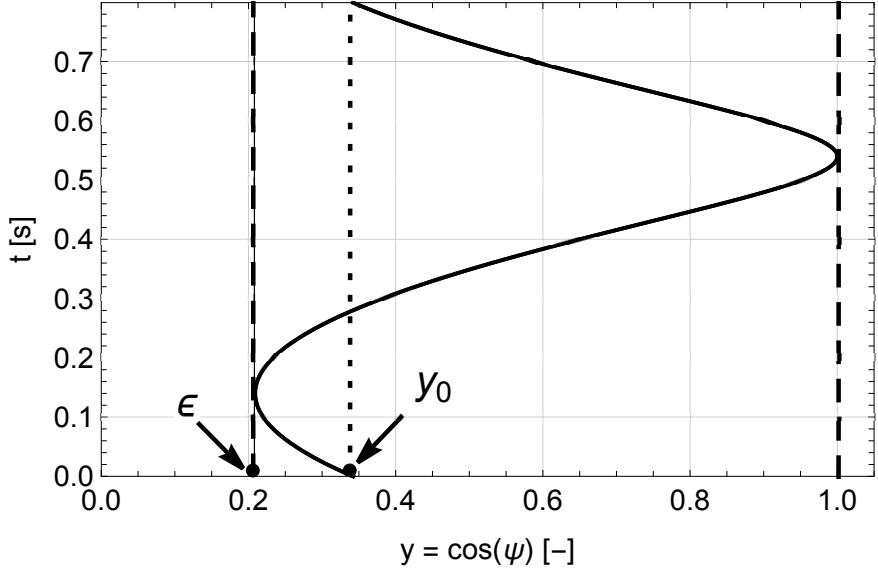}
    \caption{Time given as Lauricella function of $y$, eq.\eqref{tempo}}
    \label{fig2}
\end{figure}
Starting from its initial value of $y$ at $t_0=0$, the time line is run in the sense decided by the sign of
$ \mathrm{d}t$ arriving to its minimum value $\epsilon$ and afterwards grows to unity as already stated as characteristic of the first possible motion.

With reference to Figure~\ref{fig1}, the polar trajectory of the  bob $m_2$ is:

\begin{equation*}
    \begin{cases}
    \rho=\sqrt{l_1^2+l_2^2+2l_1l_2 \cos(\psi)}\\
    \mu=\theta+\alpha
    \end{cases}
\end{equation*}
where $\alpha$ comes from:
\begin{equation*}
    \begin{cases}
    \sin(\alpha)=l_2\sin(\psi)/\rho\\
    \cos(\alpha)=\dfrac{\rho^2+l_1^2-l_2^2}{2l_1\rho}
    \end{cases}
\end{equation*}

We assumed:

\begin{equation*}
    l_1=1 \quad \text{m}, \quad l_2=2 \quad \text{m}
\end{equation*}

\begin{figure}[H]
    \centering
    \includegraphics[scale=1]{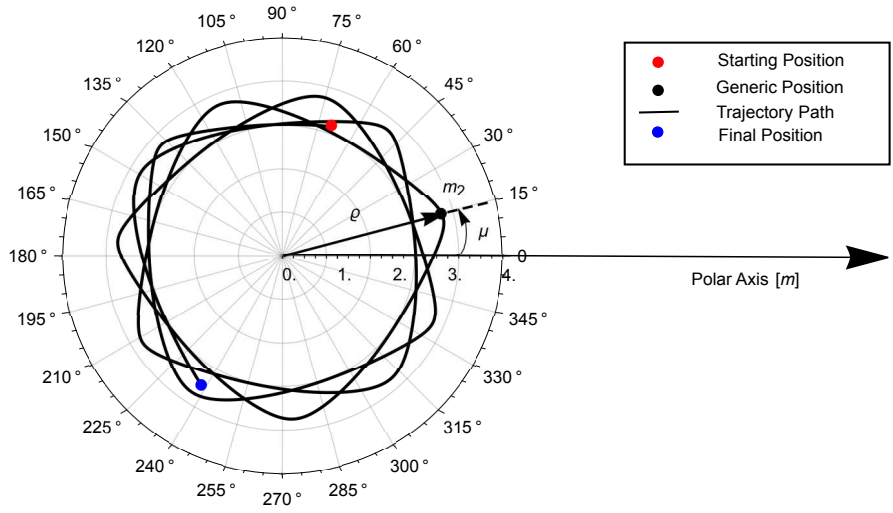}
    \caption{First motion: GFDP $(\rho,\mu) $ orbit}
    \label{fig3}
\end{figure}
Of course the above formulae for  $\rho,\mu,\alpha$ allow the construction of polar orbit for each case.
In such a way the $m_2$ -trajectory has been analytically defined, completely.
\section{Second motion:  \texorpdfstring{$-1<\epsilon<0 $ }{TEXT}}

For clarity's sake, in this and next case we will not use  the notation  $\epsilon$, but the symbol 
$-|\epsilon|$. Furthermore all the treatments will be shortened, restricting them to the main analytical things.

If $\epsilon \in [-1,0]$ we have $y\in]-|\epsilon|,1[$ and the change of variable is:
\begin{equation*}
    s=u+ \frac{1-|\epsilon|}{2}
\end{equation*}

and it centers the new reference at the half of the interval $]-|\epsilon|,1[$.

If now we define: 

\begin{equation*}
    \sigma=\frac{1-|\epsilon|}{2}, \quad  \hat{\sigma}(y)=\frac{y-\sigma}{1-\sigma}\sqrt{\frac{\beta(\delta^2-\sigma^2)}{2B(1+\sigma)}}, \quad  \bar{\sigma}=4\hat{\sigma}(1) , \quad \tilde{\sigma}(y)=\frac{A\hat{\sigma}(y)}{2\beta(\zeta+\sigma)}
\end{equation*}

then we can get:

\begin{align*}
    t-t_0= \pm \left[\hat{\sigma}(s)F_D^{(5)} \left(\left.\begin{array}{c}\begin{array}{cc}
      1;   & -\dfrac{1}{2},-\dfrac{1}{2},\dfrac{1}{2},\dfrac{1}{2},\dfrac{1}{2}\end{array}\\
      \\
       2
    \end{array} \right| \dfrac{\sigma-s}{\sigma-\delta},\dfrac{\sigma-s}{\sigma+\delta},\dfrac{s-\sigma}{1-\sigma},\dfrac{\sigma-s}{1-\sigma},\frac{\sigma-s}{1+\sigma}\right)\right]_{y_0}^y
\end{align*}

As before, we can base on IRT for computing the period through a  $F_D^{(4)}$ as:

\begin{equation*}
    T_{2} = \bar{\sigma} \sum_{i=1}^2 F_D^{(4)} \left(\left.\begin{array}{c}\begin{array}{cc}
      1;   & -\dfrac{1}{2},-\dfrac{1}{2},\dfrac{1}{2},\dfrac{1}{2}\end{array}\\
      \\
       \dfrac{3}{2}
    \end{array} \right| (-1)^i \dfrac{\sigma-1}{\sigma-\delta},(-1)^i\dfrac{\sigma-1}{\sigma+\delta},(-1)^i\dfrac{\sigma-1}{1+\sigma},-1\right) 
\end{equation*}

For $I_2(y)$ we have:

\begin{equation*}
    I_2(y)=\left[\tilde{\sigma}(s)F_D^{(6)} \left(\left.\begin{array}{c}\begin{array}{cc}
      1;   & -\dfrac{1}{2},-\dfrac{1}{2},\dfrac{1}{2},\dfrac{1}{2},\dfrac{1}{2},1\end{array}\\
      \\
       2
    \end{array} \right| \dfrac{\sigma-s}{\sigma-\delta},\dfrac{\sigma-s}{\sigma+\delta},\dfrac{s-\sigma}{1-\sigma},\dfrac{\sigma-s}{1-\sigma},\frac{\sigma-s}{1+\sigma},\dfrac{\sigma-s}{\zeta+s}\right)\right]_{y_0}^y
\end{equation*}

\subsection{Sample problem, second motion}
 The relevant choice of parameters is:
\begin{equation*}
    \nu=\frac{1}{2}, \quad \beta=\frac{1}{2}, \quad \psi_0=88 \degree, \quad \theta_0=22.5 \degree, \quad \dot{\psi}_0=2 \quad \frac{\text{rad}}{s}, \quad \dot{\theta}_0=1 \quad\frac{\text{rad}}{\text{s}}
\end{equation*}

With reference to the previous sample problem, only the $\psi_0$ value has been changed, so that  $\epsilon=-0.123205$.

\begin{figure}[H]
    \centering
    \includegraphics[scale=0.9]{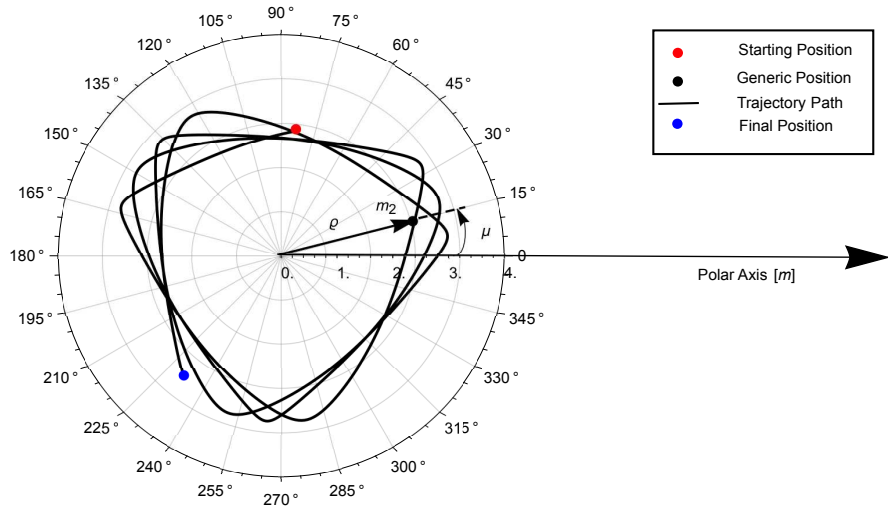}
    \caption{Second motion: GFDP $(\rho,\mu) $ orbit}
    \label{fig4}
\end{figure}

\section{Third motion: case \texorpdfstring{ $\epsilon<-1$}{TEXT}}
As it concerns the case $\epsilon<-1$, no change is necessary. As before, we get:

\begin{align*}
    t-t_0= \pm \delta \sqrt{\frac{\beta}{2B|\epsilon|}} \left[sF_D^{(5)} \left(\left.\begin{array}{c}\begin{array}{cc}
      1;   & -\dfrac{1}{2},-\dfrac{1}{2},\dfrac{1}{2},\dfrac{1}{2},\dfrac{1}{2}\end{array}\\
      \\
       2
    \end{array} \right| \dfrac{s}{\delta},-\dfrac{s}{\delta},-\dfrac{s}{|\epsilon|},-s,s\right)\right]_{y_0}^y
\end{align*}

with the period given by:

\begin{equation*}
    T_{3}=2\delta \sqrt{\frac{2\beta}{B|\epsilon|}}\sum_{i=1}^2 F_D^{(4)} \left(\left.\begin{array}{c}\begin{array}{cc}
      1;   & -\dfrac{1}{2},-\dfrac{1}{2},\dfrac{1}{2},\dfrac{1}{2}\end{array}\\
      \\
       \dfrac{3}{2}
    \end{array} \right| \dfrac{1}{\delta},-\dfrac{1}{\delta},\dfrac{(-1)^{i}}{|\epsilon|},-1\right) 
\end{equation*}

and:

\begin{equation*}
I_2(y)=
    \frac{A\delta}{2\zeta} \sqrt{\frac{1}{2\beta B|\epsilon|}} \left[sF_D^{(6)} \left(\left.\begin{array}{c}\begin{array}{cc}
      1;   & -\dfrac{1}{2},-\dfrac{1}{2},\dfrac{1}{2},\dfrac{1}{2},\dfrac{1}{2},1\end{array}\\
      \\
       2
    \end{array} \right| \dfrac{s}{\delta},-\dfrac{s}{\delta},-\dfrac{s}{|\epsilon|},-s,s,-\dfrac{s}{\zeta
    }\right)\right]_{y_0}^y
\end{equation*}

\subsection{Fourier coefficients}
First of all we have:
\begin{equation*}
    a_0=\frac{16 \delta}{3T} \sqrt{\frac{\beta}{2B|\epsilon|}} \sum_{i=1}^2 (-1)^{i+1} F_D^{(4)} \left(\left.\begin{array}{c}\begin{array}{cc}
      2;   & -\dfrac{1}{2},-\dfrac{1}{2},\dfrac{1}{2},\dfrac{1}{2}\end{array}\\
      \\
       \dfrac{5}{2}
    \end{array} \right| \dfrac{1}{\delta},-\dfrac{1}{\delta},\dfrac{(-1)^{i}}{|\epsilon|},-1\right) 
\end{equation*}
We define now:
\begin{equation*}
    w(s)=\delta \sqrt{\frac{\beta}{2B|\epsilon|}} sF_D^{(5)} \left(\left.\begin{array}{c}\begin{array}{cc}
      1;   & -\dfrac{1}{2},-\dfrac{1}{2},\dfrac{1}{2},\dfrac{1}{2},\dfrac{1}{2}\end{array}\\
      \\
       2
    \end{array} \right| \dfrac{s}{\delta},-\dfrac{s}{\delta},-\dfrac{s}{|\epsilon|},-s,s\right)
\end{equation*}

Let it be:
\begin{equation*}
    \hat{t}=w(y_0), \quad \tilde{t}=w(-1)
\end{equation*}

Considering a period and assuming $\dot{y}_0<0$, the piecewise defined function which describes the $t=t^{-1}(y)$ ad its derivative are:

\begin{equation*}
t(y)=
    \begin{cases}
    \hat{t}-w(y) & y \in [y_0,-1]\\
    \hat{t}-2\tilde{t}+w(y) & y\in[-1,1]\\
    \hat{t}+T-w(y) & y\in [1,y_0]
    \end{cases}, \quad
    t'(y)=
    \begin{cases}
    -\hat{f}(y)  & y \in [y_0,-1]\\
    \hat{f}(y) & y \in [-1,1] \\
    -\hat{f}(y) &  y\in [1,y_0]
    \end{cases}
\end{equation*}

Then, by the approach as before, we obtain:

\begin{equation*}
    \hat{a}_n=\frac{2}{\xi_n}\cos(\xi_n(\tilde{t}-\hat{t})) \int_{-1}^1 \sin(\xi_n(\tilde{t}-w(s)))ds
\end{equation*}
And:
\begin{equation*}
    \hat{b}_n=-\tan(\xi_n(\tilde{t}-\hat{t})) \hat{a}_n
\end{equation*}
Then:
\begin{equation*}
    y(t)= \frac{a_0}{2}+\frac{2}{T} \sum_{n=1}^{+\infty} \frac{\hat{a}_n}{\cos (\xi_n(\tilde{t}-\hat{t})) } \cos(\xi_n(t+\tilde{t}-\hat{t}))
\end{equation*}

\subsection{Sample problem, third motion}

\begin{equation*}
    \nu=\frac{1}{2}, \quad \beta=\frac{1}{2}, \quad \psi_0=160 \degree, \quad \theta_0=22.5 \degree, \quad \dot{\psi}_0=2 \quad \frac{\text{rad}}{s}, \quad \dot{\theta}_0=1 \quad\frac{\text{rad}}{\text{s}}
\end{equation*}

With reference to the previous sample problem, only the $\psi_0$ value has been changed, so that $\epsilon=-1.0338$.

\begin{figure}[H]
    \centering
    \includegraphics[scale=0.85]{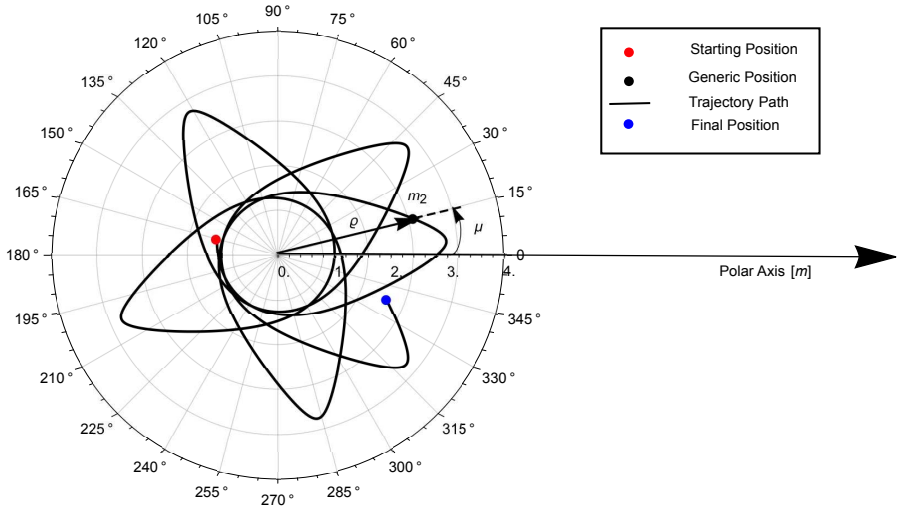}
    \caption{Third motion: GFDP $(\rho,\mu )$ orbit}
    \label{fig5}
\end{figure}

\section{Conclusions}

The  gravity free double pendulum, GFDP,  has two degrees of freedom, say the functions $\theta(t)$ and $\psi(t)$ whose detection led us to hyperelliptic integrals with the critical role of a non dimensional parameter $\epsilon$ which holds all the system data.

In such a way the GFDP will behave differently according to the $\epsilon$ spans. 

Three possible ranges have been found for it and for only one of them  the angle $\psi$ will be free of describing all values of its turn.
The a.m. integrals, thanks to the hypergeometric functions IRT, have been expressed in terms of different Lauricella functions. Such a representation is notoriously true if and only if a special constraint on the state variables is met; then, by means of a suitable change of variable, we make sure that such a condition is fully satisfied. Once the function $t=t(\psi)$ has been inverted via the Fourier series, we arrive at $\cos\psi(t)$ as a Fourier series of time. The function $\theta$ is obtained solving a further hyperelliptic integral by means of a higher degree Lauricella function holding  $\psi$  too; in such a way $\theta$ comes as function of $\psi$ and -thanks to the a.m. Fourier expansion- of time. 
These  Fourier coefficients have been computed numerically,  and mean the only one aspect of our work not expressed in analytic closed form.
All kinematic elements of the system are then computable, providing a complete knowledge of the time laws for $m_1$ and $m_2$ and then of their trajectories, Figure~\ref{fig3}, Figure~\ref{fig4}, Figure~\ref{fig5}. 

Let us analyze the $\Gamma$ polar plots relevant to the proposed sample problems, obtained by means of the closed form formulae given above and tested by the ODEs numerical treatment.

The common value selected for each of the motions is $\theta_0=\pi/8$. The starting point is red-marked while the last computed is blue. The time duration is 10 seconds for each simulation.

Unlikely to $\psi$, there is no restriction on $\theta$ so that the trajectory of the point mass $m_1$ has an angular period of $2\pi$ whilst this is not valid for $\psi$ which varies between $ \arccos (\epsilon)$ and $2\pi$ in the first motion and between $\arccos(-|\epsilon|)$ and $2\pi$ in the second.
Generally speaking, the above periods of $m_1$ (angle $\theta$) and $m_2$ (angle $\psi$) will not be commensurable each other: so that the global motion will not be periodic almost for a reasonable amount of time of simulation.

What we found in this paper is then resembling what happens in celestial mechanics. When to the basic attraction law $1/r^2$ is added the bulge effect of the attractor, the original elliptic trajectory of the attracted body is forced to rotate around the attractor, so that its resulting orbit consists of nothing but a sequence of not overlapping ellipses shifted each other, what is called "rosette".
But in our cases of motion we do not have any regular micro-orbits such those described above, being the single not overlapped loops quite irregular, so that our rosette appears to be a bit deformed. Anyway, in the third case its aspect becomes more regular when $\psi$ also is free of taking each possible value like $\theta$.

\section{Appendix: Short about Lauricella functions}

Hereinafter an outline of the Lauricella
hypergeometric functions.

The first hypergeometric historical series appeared in the Wallis's \textit{Arithmetica
infinitorum} (1656), the general expression of $_{2}F_{1}$ is:

\begin{equation*}
_2\mathrm{F}_{1}\left(\left. 
\begin{array}{c}
a,b\\[2mm]
c
\end{array}
\right|x\right)=\sum_{n=0}^{\infty }\frac{\left( a\right) _{n}\left(
b\right) _{n}}{\left( c\right) _{n}}\frac{x^{n}}{n!}, \quad |z|<1
\end{equation*}

A meaningful contribution on various $_{2}F_{1}$ topics
is ascribed to Euler in three papers \cite{euler1738progressionibus, euler1769curva, euler1845institutionum};
but he does not seem to have known the integral representation involving the $\Gamma$ function too:
\[
_2\mathrm{F}_{1}\left(\left. 
\begin{array}{c}
a,b\\[2mm]
c
\end{array}
\right|x\right)=\frac{\Gamma (c)}{\Gamma (a)\Gamma (c-a)}\,\int_{0}^{1}%
\frac{u^{a-1}(1-u)^{c-a-1}}{(1-xu)^{b}}\,\mathrm{d}u
\]
really due to Legendre \cite{legendre1816exercices}. The above integral relationship is true if $c>a>0$ and for 
$\left| x\right| <1,$ even if this limitation can be discarded thanks to the analytic
continuation.

Many functions have been introduced in 19$^{\mathrm{th}}$ century for
generalizing the hypergeometric functions to multiple variables such as the Appell ${\rm F}_1$ two-variable hypergeometric series.

The hypergeometric functions introduced and investigated by G. Lauricella (1893) \cite{lauricella1893sulle} and S. Saran (1954) \cite{saran1955transformations}, are those of our prevailing interest; and among them the function $F_{D}^{(n)}$ of $n\in \mathbb{N}^{+}$ variables (and $n+2$
parameters) defined as: 
\[
\mathrm{F}_{D}^{(n)}\left(\left. 
\begin{array}{c}
a,\bm{b}\\[2mm]
c
\end{array}
\right|\bm{x}\right):=
\sum_{m_{1},\ldots ,m_{n}\in \mathbb{N}}\frac{(a)_{m_{1}+\cdots
+m_{n}}(b_{1})_{m_{1}}\cdots (b_{n})_{m_{n}}}{(c)_{m_{1}+\cdots
+m_{n}}m_{1}!\cdots m_{n}!}\,x_{1}^{m_{1}}\cdots x_{n}^{m_{m}} 
\]
with $\bm{b}=b_1,...,b_n$, $\bm{x}=x_1,...,x_n$ and
with the hypergeometric series usual convergence requirements $%
|x_{1}|<1,\ldots ,|x_{n}|<1$. If $\mathrm{Re}\,(c)>\mathrm{Re}\,(a)>0$ , the
relevant Integral Representation Theorem provides: 
\begin{equation}\label{iirtt}
\mathrm{F}_{D}^{(n)}\left(\left. 
\begin{array}{c}
a,\bm{b}\\[2mm]
c
\end{array}
\right|\bm{x}\right)=\frac{%
\Gamma (c)}{\Gamma (a)\,\Gamma (c-a)}\,\int_{0}^{1}\,\frac{%
u^{a-1}(1-u)^{c-a-1}}{(1-x_{1}u)^{b_{1}}\cdots (1-x_{n}u)^{b_{n}}}\,\mathrm{d%
}u 
\end{equation}
allowing the analytic continuation to $\mathbb{C}^{n}$ deprived of the
Cartesian $n$-dimensional product of the interval $]1,\infty [$ with itself.

Such functions are not much pervasive in Mathematical Physics \cite{scarpello2011exact,slater1966generalized} and have their better employ in providing a valuable help in hyperelliptic integrations, being the IRT used as a tool to transfer the integration task to a hypergeometric summation.

The Lauricella functions -\textit{in lack of specific SW packages} -have been implemented in this work thanks to some reduction theorems which will form the object of a next paper. 
Here it will be enough to anticipate the following. 
All the Lauricella series are multi-valued power series whose coefficients are other series nested each other: so that e.g. $F_{D}^{(3)}$ consists of a triple infinite power series. But-by a proper work in re-scaling indices and other manipulations- it can be reduced to a single series whose coefficients are depending on a convenient $_{2}F_{1}$ function \textit{for which SW packages are by long time available}.
Furthermore $F_{D}^{(5)}$ has been led to a simple power series whose coefficients depend on 3 computable blocks each of them holds again the a.m. Gauss hypergeometric $_{2}F_{1}$.
All above really helps such functions management increasing the series convergence speed.

\newpage

\noindent{\Large\bf Competing Interests}\\\\
Authors declare that no conflicts of interest  exist.\\[3mm]

\scriptsize\-----------------------------------------------------------------------------------------------------------------------------------------\\\copyright \it 20YY  Author name; This is an Open Access article distributed under the terms of the Creative Commons Attribution License
\href{http://creativecommons.org/licenses/by/2.0}{http://creativecommons.org/licenses/by/4.0},  which permits unrestricted use, distribution, and reproduction in any medium,
provided the original work is properly cited.

\end{document}